\documentclass[aps,prb,twocolumn,floatfix,superscriptaddress,showpacs,psfig]{revtex4-2}

\usepackage[colorlinks=true,urlcolor=blue,citecolor=blue,linkcolor=blue]{hyperref}
\usepackage{graphicx}
\usepackage{color}
\usepackage{epsfig}
\usepackage{amsmath}
\usepackage{amssymb}
\usepackage{mathrsfs}
\usepackage{bm}
\usepackage{subfigure}
\usepackage{float}
\usepackage{url}
\usepackage{enumerate}
\usepackage{braket}
\usepackage{comment}
\usepackage{multirow}
\usepackage{appendix}
\usepackage[section]{placeins}

\definecolor{violet}{rgb}{0.56,0.0,1.0}

\begin{document}

\title{Ground-state phase diagram and route to supersolidity in a two-component extended Bose–Hubbard model}

\author{Xuexin Qiu}
\affiliation{School of Physical Science and Technology $\&$ Key Laboratory of Quantum Theory and Applications of MoE, Lanzhou University, Lanzhou 730000, China.}
\affiliation{Lanzhou Center for Theoretical Physics, Key Laboratory of Theoretical Physics of Gansu Province, Lanzhou University, Lanzhou 730000, China.}

\author{Ning Bian}
\affiliation{School of Physical Science and Technology $\&$ Key Laboratory of Quantum Theory and Applications of MoE, Lanzhou University, Lanzhou 730000, China.}
\affiliation{Lanzhou Center for Theoretical Physics, Key Laboratory of Theoretical Physics of Gansu Province, Lanzhou University, Lanzhou 730000, China.}

\author{Yang Liu}
\affiliation{School of Physical Science and Technology $\&$ Key Laboratory of Quantum Theory and Applications of MoE, Lanzhou University, Lanzhou 730000, China.}
\affiliation{Lanzhou Center for Theoretical Physics, Key Laboratory of Theoretical Physics of Gansu Province, Lanzhou University, Lanzhou 730000, China.}

\author{Saisai He}
\affiliation{School of Physical Science and Technology $\&$ Key Laboratory of Quantum Theory and Applications of MoE, Lanzhou University, Lanzhou 730000, China.}
\affiliation{Lanzhou Center for Theoretical Physics, Key Laboratory of Theoretical Physics of Gansu Province, Lanzhou University, Lanzhou 730000, China.}

\author{Z. Y. Xie}
\email{qingtaoxie@ruc.edu.cn}
\affiliation{School of Physics, Renmin University of China, Beijing 100872, China}
\affiliation{Key Laboratory of Quantum State Construction and Manipulation of MoE, Renmin University of China, Beijing 100872, China}

\author{Hong-Gang Luo}
\affiliation{School of Physical Science and Technology $\&$ Key Laboratory of Quantum Theory and Applications of MoE, Lanzhou University, Lanzhou 730000, China.}
\affiliation{Lanzhou Center for Theoretical Physics, Key Laboratory of Theoretical Physics of Gansu Province, Lanzhou University, Lanzhou 730000, China.}

\author{Jize Zhao}
\email{zhaojz@lzu.edu.cn}
\affiliation{School of Physical Science and Technology $\&$ Key Laboratory of Quantum Theory and Applications of MoE, Lanzhou University, Lanzhou 730000, China.}
\affiliation{Lanzhou Center for Theoretical Physics, Key Laboratory of Theoretical Physics of Gansu Province, Lanzhou University, Lanzhou 730000, China.}

\begin{abstract}
	We investigate the ground-state phase diagram of a two-component extended Bose–Hubbard model recently realized with dipolar excitons, using infinite projected entangled-pair states. For the experimentally relevant parameter regime, checkerboard, Mott-insulating, superfluid, and vacuum phases are identified. These phases exhibit orbital-selective character, wherein the two components occupy different quantum states, but we find no evidence for a supersolid phase. The absence of supersolidity is attributed to the strongly interaction-dominated microscopic energy scales, which severely restrict the superfluid regime. Guided by the supersolid mechanism, we further explore a nearby parameter regime with enhanced hopping of one component and identify an orbital-selective supersolid phase. Further finite bond-dimension and unit-cell analyses establish the robustness of this phase. 
Our results clarify the zero-temperature phase structure of the dipolar-exciton platform and provide a possible route toward realizing supersolidity in this setting.
\end{abstract}

\maketitle

\section{Introduction}
Similar to the Hubbard model for interacting lattice Fermi systems, the interacting bosons on a lattice can be described by the so-called Bose-Hubbard~(BH) model. This minimal model 
provides a framework to investigate the competition between kinetic energy and on-site repulsive interactions. 
The hallmark achievement of the BH model is its successful prediction of the quantum phase transition between the superfluid~(SF) phase and the Mott-insulating~(MI) phase~\cite{PhysRevB.40.546, PhysRevB.44.10328, PhysRevLett.81.3108}, which was observed by later experiments~\cite{Greiner2002, Bakr2009, doi:10.1126/science.1192368, Weitenberg2011}.
While the standard BH model has been extraordinarily successful, its limitation is also obvious in that it only considers the on-site interaction term. In many physically relevant 
systems, interactions can be significant over finite distances. 
To capture these effects, one must generalize the model to the extended Bose-Hubbard~(EBH) model, which incorporates nearest-neighbor (and sometimes longer-range) density-density 
interactions, providing a minimal framework for understanding how kinetic delocalization competes with interaction-driven density ordering. 
In its single-component form, this competition gives rise to SF, MI, density-wave, and, under suitable conditions, supersolid~(SS) 
phases~\cite{PhysRevLett.84.1599, PhysRevLett.88.170406, PhysRevLett.89.130401, PhysRevLett.94.207202, PhysRevB.75.094501, doi:10.1126/science.aac9812, doi:10.1126/science.adq7082}.
Among these, supersolidity has attracted long-standing interest~\cite{RevModPhys.84.759, Leonard2017, PhysRevLett.122.130405, PhysRevX.9.011051, PhysRevX.9.021012, Norcia2021} because 
it requires the coexistence of off-diagonal phase coherence and density modulation, two tendencies that prefer suppressing each other. Thus, realizing supersolidity 
requires a rather delicate balance between hopping and intersite repulsion.

Two-component bosonic systems are known to host even richer physics, such as counterflow superfluidity and paired phases~\cite{PhysRevLett.90.100401, Altman_2003}.
Recent experiments~\cite{Lagoin2022} opened a new setting for exploring this problem by realizing a two-component EBH model with dipolar excitons confined in an artificial 
square lattice. In contrast to conventional single-component EBH systems, this platform involves two inequivalent Wannier orbitals with distinct hopping amplitudes and interaction scales. 
This naturally introduces richer competing tendencies and allows orbital-selective~\cite{Anisimov2002} phases in which the two components may occupy different quantum states. Such a multicomponent 
setting raises an important question: what is the ground-state phase diagram of this model, and can it host an SS phase?

Previous studies~\cite{Lagoin2022} based on mean-field theory and exact diagonalization have identified checkerboard~(CB), MI and normal fluid behavior in this system and provided a preliminary picture of its finite-temperature properties. However, the ground-state phase diagram remains less understood, particularly regarding the possible existence of supersolidity and the role of quantum correlations beyond mean-field treatments. Since the experimentally relevant regime involves strong interactions and competing orders, a controlled many-body treatment is necessary to address these questions.

In this work, we investigate the ground-state phase diagram of the two-component EBH model using infinite projected entangled-pair states~(iPEPS)~\cite{2004cond.mat7066V, RevModPhys.93.045003}. 
For the experimental parameter set, we find no SS phase. Instead, the system hosts CB, MI, vacuum~(VM) phases, and two small SF sectors. We further show that supersolidity can be stabilized by increasing the hopping in one of the two components, from which a robust orbital-selective CB-SS phase emerges. These results clarify the ground-state physics of the dipolar-exciton platform and provide a concrete route toward realizing supersolidity in this setting.

The paper is organized as follows. In Sec.~\ref{sec:model}, we introduce the model, the iPEPS ansatz, and the numerical method. Section~\ref{sec:result} presents the phase diagrams, representative cuts, and consistency checks, with particular emphasis on the robustness of the CB-SS phase. We summarize the main conclusions in Sec.~\ref{sec:summary}.

\section{Model and Method}
\label{sec:model}
Following Ref.~\cite{Lagoin2022}, the two-component EBH Hamiltonian is written as
\begin{equation}
H=\sum_i h_i + \sum_{\langle i,j \rangle} h_{ij}.
\label{H}
\end{equation}
Here the onsite term $h_i$ is
\begin{equation}
h_i=\sum_{\alpha,\beta,\gamma,\delta}U_{\alpha\beta\gamma\delta}b^{\dagger}_{i\alpha}b^{\dagger}_{i\beta}b_{i\gamma}b_{i\delta}-\sum_{\alpha}\mu_{\alpha}n_{i\alpha},
\label{hi}
\end{equation}
and the nearest-neighbor term $h_{ij}$ is
\begin{equation}
h_{ij}=-\sum_{\alpha}t_{\alpha}(b^{\dagger}_{i\alpha}b_{j\alpha}+{\rm h.c.})+\sum_{\alpha,\beta,\gamma,\delta}V_{\alpha\beta\gamma\delta}b^{\dagger}_{i\alpha}b^{\dagger}_{j\beta}b_{j\gamma}b_{i\delta}.
\label{hij}
\end{equation}
In Eqs.~(\ref{hi}) and (\ref{hij}), $\alpha$, $\beta$, $\gamma$, and $\delta$ take values 1 or 2, labeling the two Wannier orbitals. The operator $b^{\dagger}_{i\alpha}$ ($b_{i\alpha}$) creates (annihilates) a boson of component $\alpha$ on site $i$, and the chemical potentials $\mu_{\alpha}$ control the filling of the $\alpha$ component.

Keeping only the dominant interaction channels identified in Ref.~\cite{Lagoin2022}, we work with the simplified Hamiltonian
\begin{align}
h_i &= U_1 n_{i1}^2 + U_2 n_{i2}^2 + U_3 n_{i1}n_{i2} - \sum_{\alpha}\mu_{\alpha}n_{i\alpha},
\label{hi2}\\
h_{ij} &= -\sum_{\alpha}t_{\alpha}(b^{\dagger}_{i\alpha}b_{j\alpha}+{\rm h.c.}) + V_1 n_{i1}n_{j1}+V_2 n_{i2}n_{j2}  \nonumber\\
	&\quad + V_{12} n_{i1}n_{j2} + V_{21} n_{i2}n_{j1}.
\label{hij2}
\end{align}
Here $U_1$, $U_2$, $U_3$ are shorthand for $U_{1111}$, $U_{2222}$, and $U_{1212}+U_{2121}+U_{1221}+U_{2112}$, and $V_1$, $V_2$, $V_{12}$ and $V_{21}$ are for $V_{1111}$, $V_{2222}$, $V_{1221}$ and $V_{2112}$, respectively. Terms proportional to $n_{i\alpha}$ are absorbed into the chemical potentials. The relevant parameters taken from Ref.~\cite{Lagoin2022} are
\begin{align}
t_1&=1, \;\: t_2=7, \;\: U_1=1000, \;\: U_2=500, \;\:  U_3=200, \nonumber\\
V_1&=35, \;\: V_2=250, \;\:  V_{12}=20, \;\:  V_{21} = 20,
\label{eq:paraset}
\end{align}
which are all in units of $\mu$eV. Because $U_1$ and $U_2$ are much larger than the remaining energy scales, it is safe in practice to treat both components as hard-core bosons, i.e., $n_{i\alpha}=0,1$.

\begin{figure}[t]
    \centering
    \includegraphics[width=0.6\columnwidth,clip]{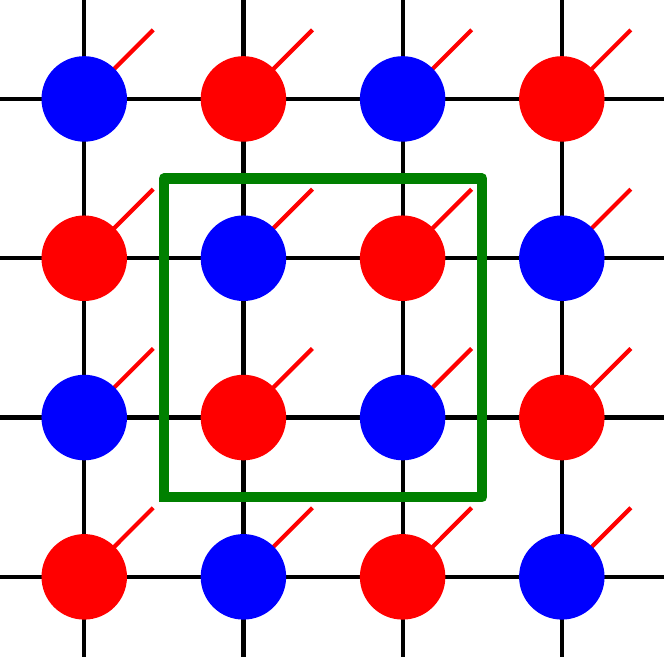}
    \caption{Schematic illustration of the iPEPS ansatz used in this work. The circles denote local tensors on the square lattice. Blue and red indicate different sublattices, red legs denote physical indices, black legs denote virtual indices, and the green square highlights a $2\times2$ unit cell.}
    \label{fig:PEPS}
\end{figure}

To study the ground-state phase diagram as a function of $\mu_1$ and $\mu_2$, we employ the iPEPS ansatz~\cite{PhysRevLett.101.250602,RevModPhys.93.045003}, a two-dimensional generalization of the matrix-product state~\cite{SCHOLLWOCK201196}. As sketched in Fig.~\ref{fig:PEPS}, the variational state is written as
\begin{align}
|\Psi\rangle &= \sum_{\{n_{xy,1},n_{xy,2}\}}
\mathrm{Tr}\!\left[\prod_{(x,y)} T^{(x,y)}_{i_{xy}j_{xy}k_{xy}l_{xy}}[n_{xy,1},n_{xy,2}]\right] \nonumber\\
&\quad\times |\ldots,n_{xy,1},n_{xy,2},\ldots\rangle,
\label{eq:PEPS}
\end{align}
where $D$, the largest value of the virtual indices~(i.e., $i_{xy}$, $j_{xy}$, $k_{xy}$ and $l_{xy}$), controls both the variational accuracy and the computational cost. To capture checkerboard order in the simplest compatible manner, our primary calculations use a $2\times2$ unit cell, while $4\times4$ calculations are performed as consistency checks at representative points.

The local tensors are optimized by imaginary-time evolution with a simple-update scheme~\cite{PhysRevLett.101.090603, Daley_2004, PhysRevLett.98.070201, PhysRevLett.101.250602}. After optimization, physical observables are evaluated using the corner transfer-matrix renormalization group (CTMRG) method~\cite{doi:10.1143/JPSJ.65.891,PhysRevB.80.094403,PhysRevB.98.235148,PhysRevB.98.235148,PhysRevLett.113.046402}. The environment dimension $\chi$ is chosen to be at least $D^2$, which is sufficient for converged expectation values in the parameter ranges considered here.

To classify the phases, we compute the average density,
\begin{equation}
\rho_{\alpha}=\frac{1}{N}\sum_i \langle n_{i\alpha} \rangle,
\end{equation}
the density-modulation order parameter,
\begin{equation}
\Delta_\alpha=\frac{1}{N}\sum_i \left|\langle n_{i\alpha}\rangle-\rho_{\alpha}\right|,
\end{equation}
and the condensate order parameter,
\begin{equation}
\psi_{\alpha}=\frac{1}{N}\sum_i \left|\langle b_{i\alpha}\rangle\right|.
\end{equation}
Here, the sum runs over all the $N$ sites in one unit cell. A finite $\psi_{\alpha}$ signals spontaneous breaking of the global $U(1)$ symmetry, suggesting that 
the system may reside in either an SF phase or SS phase.
They can be further distinguished by checking the lattice translational symmetry, which is signalled by $\Delta_\alpha$. 
On the other hand, when $\psi_\alpha=0$, the system is likely in the density-wave state, MI state or VM state, which can be further detected 
by $\Delta_\alpha$ and $\rho_\alpha$.

\section{Results and Discussion}
\label{sec:result}
In this section we will present our numerical results, which are organized around two central questions. The first is why the supersolidity is absent 
in the experimental parameter regime and the second is how such a phase can be stabilized by tuning the microscopic parameters.

\subsection{Experimental parameter regime}

\begin{figure}[t]
    \centering
    \includegraphics[width=0.9\columnwidth,clip]{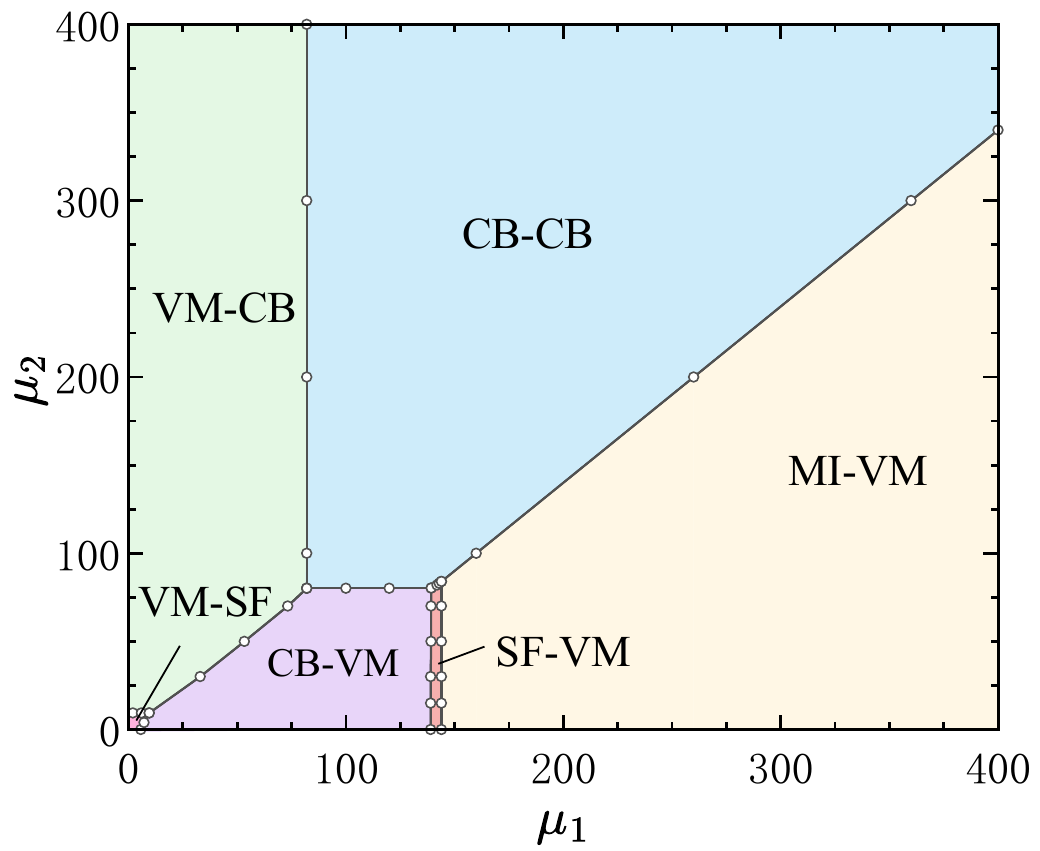}\\[-2pt]
        \caption{Ground-state phase diagram for the parameter set identified in the experiment~\cite{Lagoin2022}, obtained by TN method with $D=4$.
	Open circles denote the transition points with the error bar smaller than the symbol size; solid lines are guides to the eye.}
    \label{fig:phasediag}
\end{figure}
We begin with the parameter set given in (\ref{eq:paraset}). This parameter set was reported in Ref.~\cite{Lagoin2022}, where
a finite-temperature phase diagram was mapped out by mean-field theory, which includes CB, MI and normal fluid phase. 
It was conjectured that an SS phase may emerge around zero temperature. In this work, we calculate the ground-state phase diagram by the tensor-network method. 
In practice, the phase diagram is plotted in the $(\mu_1,\mu_2)$ plane by first performing a coarse scan to identify candidate phase boundaries, followed by refined scans 
with dense cuts in $\mu_1$ at fixed $\mu_2$. This refinement is necessary because compressible regions are much narrower than the dominant incompressible sectors.
\begin{figure*}[btp]
  \centering
  \begin{minipage}[]{0.45\textwidth}
    \centering
    \includegraphics[width=\linewidth]{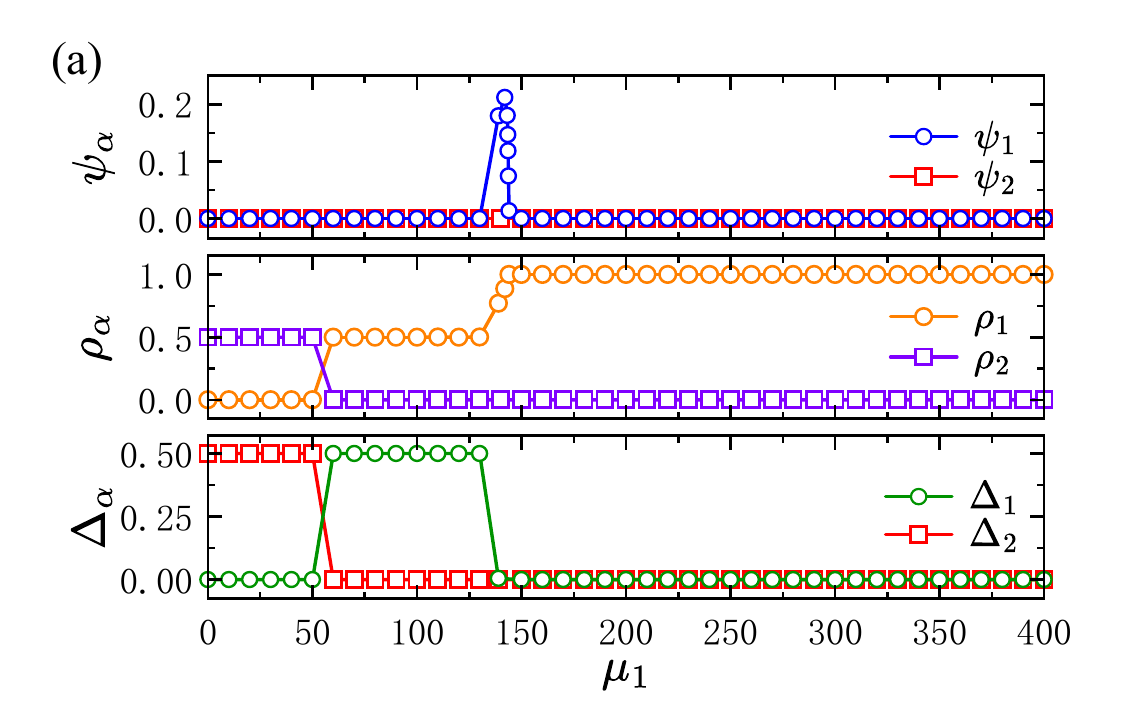}
  \end{minipage}
  \hfill
  \begin{minipage}[]{0.45\textwidth}
    \centering
    \includegraphics[width=\linewidth]{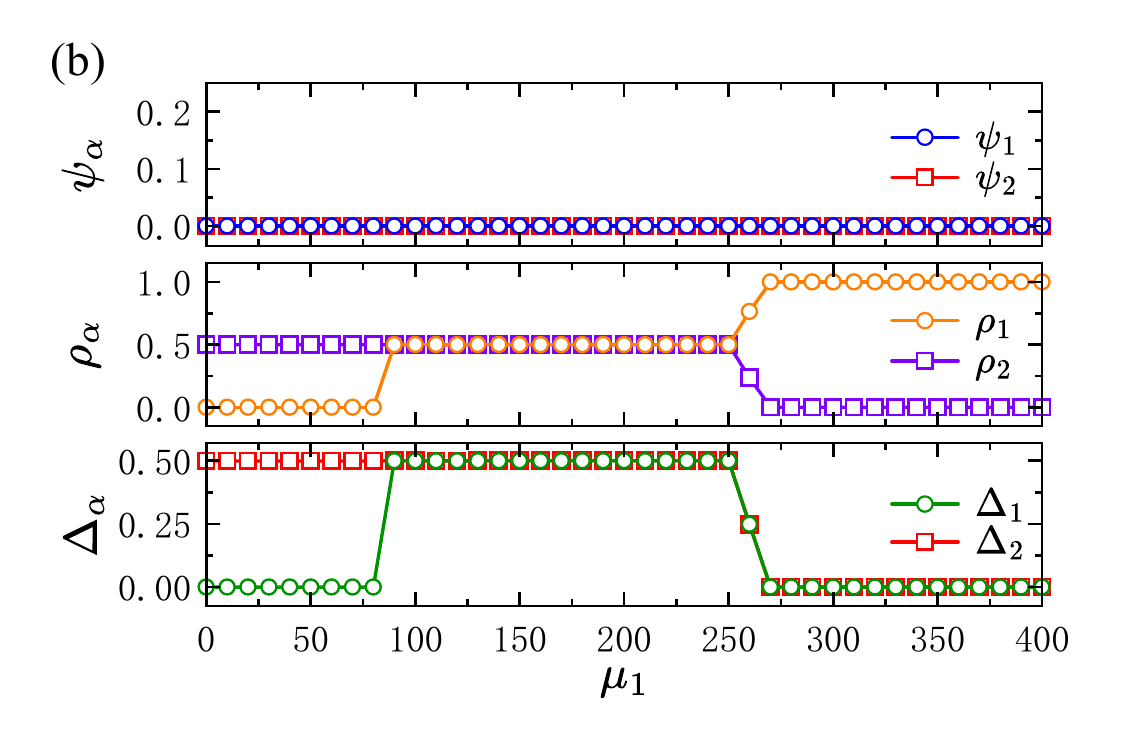}
  \end{minipage}
\caption{Order parameters $\psi_1$, $\psi_2$, $\rho_1$, $\rho_2$, $\Delta_1$ and $\Delta_2$ for the experimental parameter set are plotted as functions of $\mu_1$ for (a) $\mu_2=50$ and (b) $\mu_2=200$.
         The discontinuities in the order parameters indicate the phase transitions are all first order. Data are obtained with $D=4$.}
\label{fig:OP}
\end{figure*}

The resulting phase diagram is shown in Fig.~\ref{fig:phasediag}. 
We identify six phases, which are marked by VM-SF, VM-CB, CB-VM, CB-CB, SF-VM and MI-VM. The first and second labels indicate the states of components 1 and 2, respectively. 
The phase diagram is dominated by incompressible states, including CB, MI and VM phases, 
with only one small SF pocket in component 2 and one narrow SF corridor in component 1. Importantly, this model is not merely a doubled copy of the single-component EBH problem~\cite{PhysRevB.65.014513}.
Orbital-selective~\cite{Anisimov2002} phases arise naturally due to the inequivalence of the two components, leading to regimes such as CB-VM and VM-CB in which the two components occupy different quantum states. These orbital-selective behaviors can be readily identified by the order parameters. For example, in 
the CB-VM phase, component 1 forms a CB density-wave state with $\psi_1\approx 0.0$ and  $\Delta_1 > 0$, while component 2
is in the VM state with $\psi_2\approx 0$, $\Delta_2\approx 0$, and $\rho_2\approx 0$.

A key feature of this parameter regime is the extreme suppression of superfluidity. The superfluid sectors of component 1 and component 2 occupy only a tiny fraction of the $(\mu_1,\mu_2)$ plane. This is the direct consequence of the fact that the hopping is too weak to compete with the interaction-driven tendency toward localization. As a result, there is no parameter window in which density modulation and phase coherence can coexist, precluding the formation of a supersolid phase. This negative result is itself physically informative. It shows that the experimentally realized two-component model sits on the strongly localized side of the competition and must be moved appreciably in model parameter space before a zero-temperature supersolid can be stabilized.

To gain an intuitive understanding of the physical origin of these phases, we examine two representative cuts through the phase diagram, as shown in Fig.~\ref{fig:OP}. 
For the first cut~(Fig.~\ref{fig:OP}(a)) at fixed $\mu_2=50$, the phase evolution with increasing $\mu_1$ is mainly governed by the competition between 
the chemical-potential imbalance and the strong intercomponent repulsion. At small $\mu_1 \ll \mu_2$, component 2 is favored and forms 
a checkerboard density-wave state stabilized by the nearest-neighbor repulsion, 
while component 1 remains depleted, giving the VM-CB phase. As $\mu_1$ increases, component 1 becomes favorable, and the filling preference switches from component 2 to component 1. 
Because the onsite intercomponent repulsion $U_3$ penalizes simultaneous occupation of the two components, component 2 is expelled and component 1 takes over the checkerboard order, leading to the CB-VM phase. 
Upon further increasing $\mu_1$, the chemical-potential gain of component 1 eventually overcomes the nearest-neighbor repulsion and drives the system toward unit filling, resulting in an MI-VM phase. 
The SF-VM sliver appears only near the crossover from the half-filled checkerboard state to the unit-filled Mott insulating state.
It spans only a narrow region $\Delta\mu_1\approx 5$~(from $\mu_1\approx 140$ to 145), within which $\psi_1$ reaches a peak value of $\sim 0.21$ while $\Delta_1$ drops to zero, 
confirming that the off-diagonal coherence is not accompanied by density modulation.  

For the second cut~(Fig.~\ref{fig:OP}(b)) at fixed $\mu_2=200$, the larger chemical potential of component 2 changes the above sequence. Since component 2 remains energetically favorable over a wider range of $\mu_1$, the system does not immediately switch from VM-CB to CB-VM. Instead, when component 1 starts to be populated, component 2 can still maintain its checkerboard modulation, and the two components coexist with density-wave order, giving rise to the CB-CB phase. Only when $\mu_1$ becomes sufficiently large does component 1 approach the Mott-insulating state, while the strong intercomponent repulsion suppresses component 2 and drives the system into MI-VM. These two cuts therefore show that the orbital-selective phase boundaries are determined not only by the chemical potentials, but also by the interplay among nearest-neighbor density ordering, intercomponent repulsion, and weak hoppings.

\subsection{Route to supersolidity in model space}
\begin{figure}[tb]
    \centering
    \includegraphics[width=0.9\columnwidth,clip]{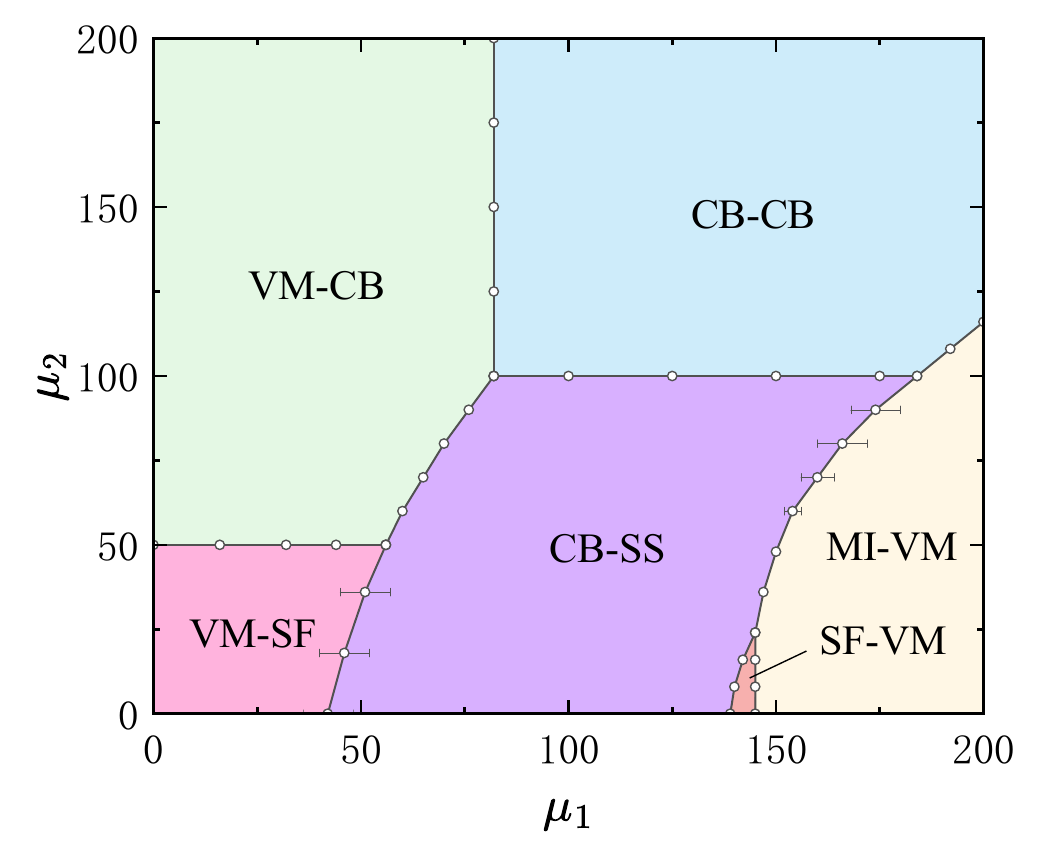}\\[-2pt]
    \caption{Ground-state phase diagram after tuning the parameter $t_2=35$, obtained by TN method with $D=4$. 
	Open circles denote the transition points; solid lines are guides to the eye. The error bars are $\pm{2}$ if not plotted explicitly.}
    \label{fig:phasediag_tuned}
\end{figure}
A natural question is whether supersolidity can be stabilized in this model. Guided by the conclusion of the one-dimensional counterpart~\cite{PhysRevB.111.235110}, 
we move along a simple path in model space and increase the hopping of component 2 to $t_2=35$, 
while keeping the remaining parameters in Eq.~(\ref{eq:paraset}) unchanged. The resulting phase diagram is shown in Fig.~\ref{fig:phasediag_tuned}. 
In addition to the VM-SF, VM-CB, CB-CB, SF-VM and MI-VM sectors inherited from the original regime, the tuned system now supports a CB-SS phase. 

The CB-SS phase is the central result in the tuned parameter regime. In this state, component 2 simultaneously exhibits a finite condensate amplitude $\psi_2$ and a finite checkerboard modulation $\Delta_2$, 
while component 1 is in CB state. Physically, the stronger hopping restores sufficient phase coherence for component 2, whereas the nearest-neighbor repulsion keeps the density modulation from melting. 
The resulting CB-SS region therefore provides a concrete and internally consistent route to supersolidity in the two-component model. 

\begin{figure}[tbp]
    \includegraphics[width=0.95\linewidth]{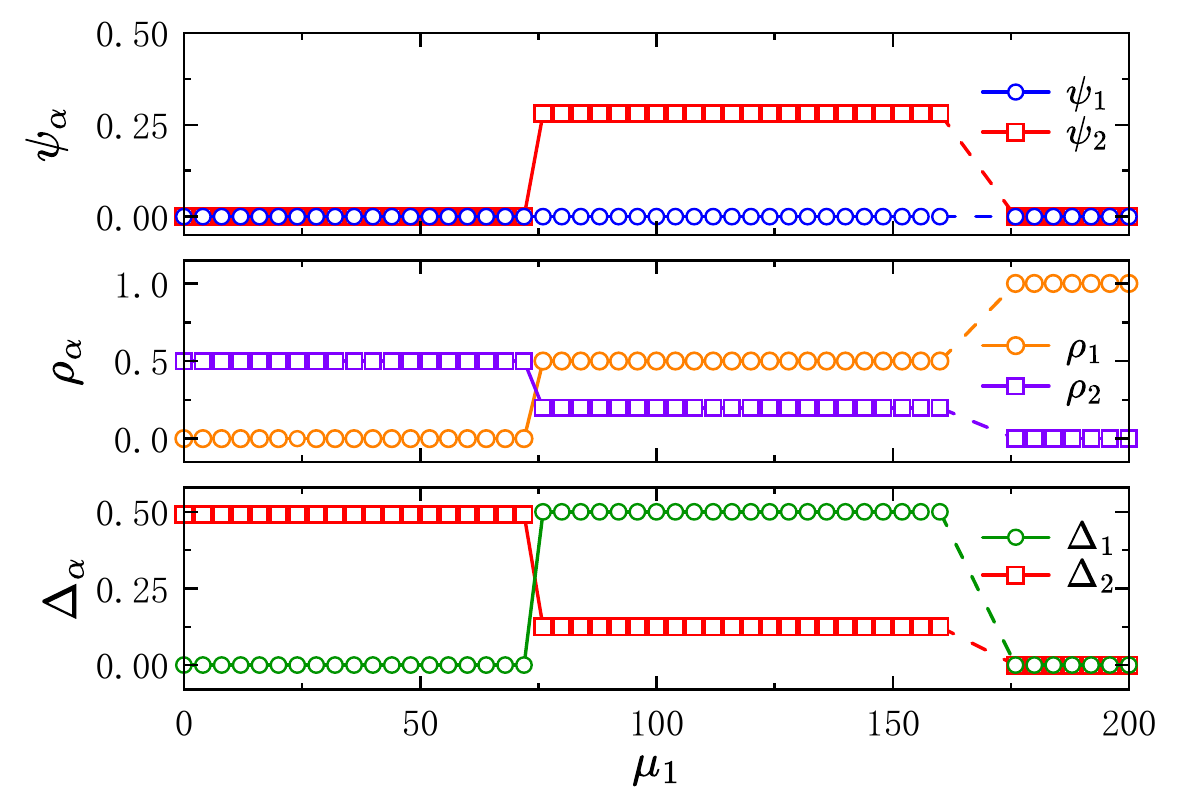}
\caption{Order parameters $\psi_1$, $\psi_2$, $\rho_1$, $\rho_2$, $\Delta_1$ and $\Delta_2$ for the tuned parameter set are plotted as functions of $\mu_1$. The discontinuities suggest that the phase transitions are first order. The dashed lines correspond to the error bars in the phase diagram. Data are obtained by TN at $\mu_2=80$ with $D=4$.}
\label{fig:OP_tuned}
\end{figure}
To further illustrate how the CB-SS phase emerges, we show in Fig.~\ref{fig:OP_tuned} the order parameters along one representative cut of the tuned phase diagram, at $\mu_2=80$.
The cut traverses three phases. At small $\mu_1$, component 2 is favored with $\rho_2 = 0.5000(0)$. The nearest-neighbor repulsion forces these particles into a CB density-wave pattern.
Meanwhile, the strong intercomponent repulsion excludes the occupation of component 1, leading to the VM-CB phase.
As $\mu_1$ increases, component 1 becomes competitive with component 2, eventually leading to a phase transition into the CB-SS phase at $\mu_1\approx 75$. 
In the CB-SS phase, the density of component 2 is suppressed due to the repulsion from component 1 but remains finite. The dilution enables component 2 to overcome the onsite repulsion  
and develop the off-diagonal coherence. Upon further increasing $\mu_1$, component 1 becomes Mott insulating and component 2 is completely expelled, resulting in an MI-VM phase. 

\begin{figure}[tb]
    \centering
    \includegraphics[width=0.95\columnwidth,clip]{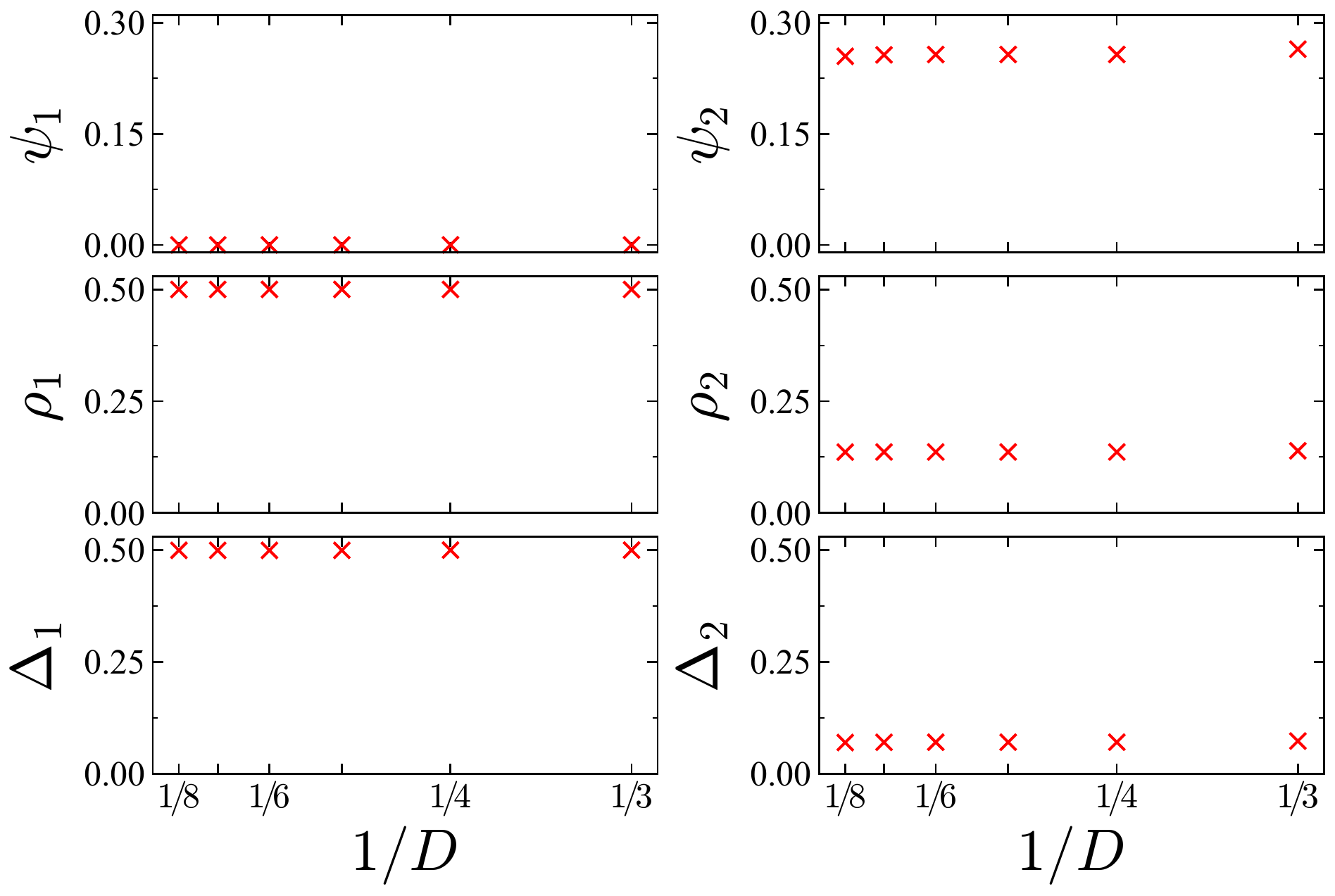}
	\caption{$1/D$ dependence ($D$=3--8) of the order parameters at the representative point~($\mu_1=120, \mu_2=50$) in the CB-SS phase.}
    \label{fig:scaling2}
\end{figure}

\begin{figure}[tb]
    \centering
    \includegraphics[width=0.95\columnwidth,clip]{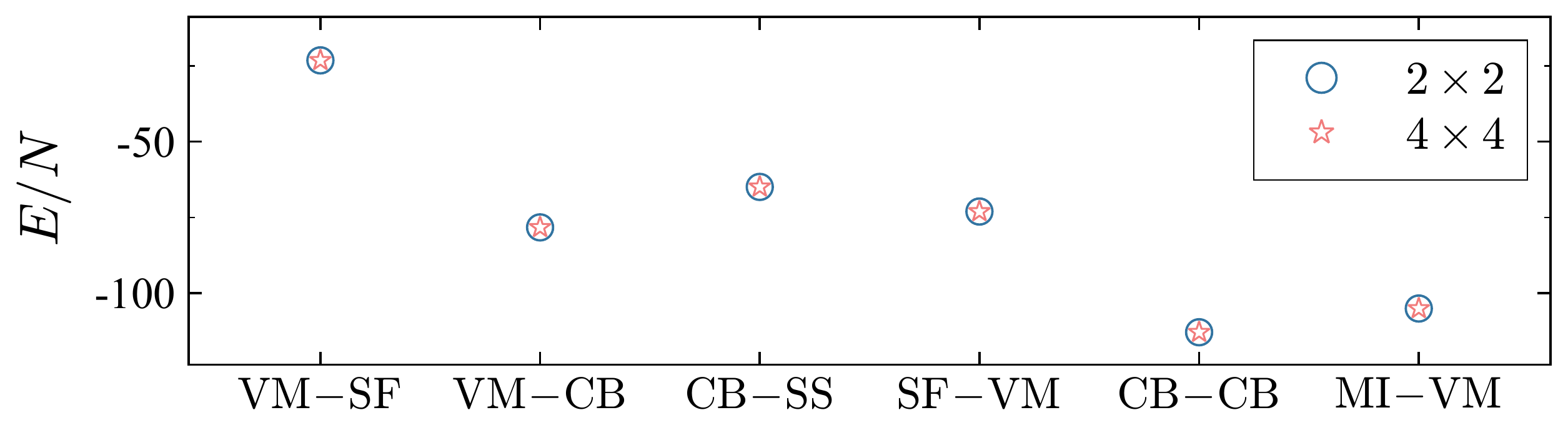}
	\caption{Ground-state energies obtained with $2\times2$ and $4\times4$ unit cells at representative points, which are $(\mu_1,\mu_2)=(25,25)$, $(50,150)$, $(120,50)$, $(143,8)$, $(150,150)$, and $(175,50)$ for the VM-SF, VM-CB, CB-SS, SF-VM, CB-CB, and MI-VM phases, respectively.}
    \label{fig:UnitCell2}
\end{figure}
The numerical stability of the CB-SS phase is supported by the consistency checks in Fig.~\ref{fig:scaling2} and Fig.~\ref{fig:UnitCell2}.
For the representative point~($\mu_1=120$, $\mu_2=50$) in the CB-SS phase [Fig.~\ref{fig:scaling2}], we plot the order parameters as a function of $1/D$.
These order parameters are smooth functions of $1/D$, which are extrapolated to the large $D$ limit by $O=O_0+O_1/D+O_2/D^2$. 
$\rho_1$ and $\Delta_1$ approach a finite value, i.e., $\rho_1=0.5000(0)$, $\Delta_1=0.4987(8)$ in the large $D$ limit while $\psi_1$ is zero. This is the character of CB phase.
On the other hand, $\psi_2$, $\rho_2$ and $\Delta_2$ converge to finite values, with $\psi_2=0.260(6)$, $\rho_2=0.1396(16)$, and $\Delta_2=0.0714(20)$, suggesting component-2 bosons are in an SS phase.
Moreover, the energies obtained from $2\times2$ and $4\times4$ unit cells~(Fig.~\ref{fig:UnitCell2}) are nearly identical. These checks show that CB-SS is neither a small $D$ artifact nor a consequence
of forcing the system into an incompatible unit cell. Instead, it is a stable phase of the tuned Hamiltonian. 

It is instructive to compare the tuned phase diagram (Fig.~\ref{fig:phasediag_tuned}) 
with the experimental one (Fig.~\ref{fig:phasediag}), which reveals how the enhanced 
hopping reshapes the phase structure. The CB-VM sector of the experimental regime is 
replaced by the CB-SS phase: the checkerboard order of component 1 is left intact, 
while component 2, instead of being completely expelled, now retains a finite density 
and condenses on top of this density-ordered background.  
In this case, increasing $t_2$ restores the phase coherence of component 2 without destroying the underlying density 
order. This is precisely the balance required for supersolidity, and it identifies the 
hopping asymmetry as the key control knob of this platform.

\section{Conclusion}
\label{sec:summary}
In this work, we have investigated the ground-state phase diagram of the two-component extended Bose–Hubbard model motivated by recent dipolar-exciton experiments, using the iPEPS tensor-network approach. For the experimentally identified parameter regime, we find vacuum, checkerboard, Mott-insulating, and two small superfluid sectors, 
which exhibit orbital-selective feature, i.e., the two components can occupy distinct quantum states. Notably, we find no evidence for a supersolid ground state. 
This absence can be understood as a direct consequence of the interaction-dominated microscopic energy scales, which suppress the coexistence of density order and phase coherence required for supersolidity.

We further identify a route toward supersolidity by tuning the hopping of one component. In this regime, a robust orbital-selective supersolid phase 
emerges. This phase remains stable under both bond-dimension scaling and enlarged-unit-cell consistency checks, establishing it as a genuine ground-state phase of the tuned Hamiltonian. 

Our results clarify the zero-temperature phase structure of the dipolar-exciton platform and demonstrate how multi-orbital competition can be exploited to stabilize supersolidity. More broadly, they provide a possible microscopic direction for future experimental efforts and suggest that multicomponent bosonic platforms may offer fertile ground for realizing and exploring unconventional supersolid phases.

\section*{Acknowledgments}
This work is supported by the National Key R$\&$D Program of China (Grants Nos. 2022YFA1402704, 2024YFA1408604, 2023YFA1406500), by the National Natural Science Foundation of China (Grants Nos. 12274187, 12247101, 12274458).

\section*{Data availability}
Numerical data underlying the figures and supporting the findings of this work are available from the corresponding author upon reasonable request.

\bibliography{main}

\end{document}